\documentclass[preprint,showpacs,preprintnumbers,amsmath,amssymb]{revtex4}

\usepackage{graphicx}
\usepackage{dcolumn}
\usepackage{bm}

\usepackage{epstopdf}
\usepackage{amsmath}

\begin{document}

\preprint{arXiv/cond-mat}

\title{Ultrafast Electron-Phonon Decoupling in Graphite}

\author{Kunie Ishioka}
\email{ishioka.kunie@nims.go.jp}
\affiliation{Advanced Nano-Characterization Center, National Institute for Materials
Science, Tsukuba, 305-0047 Japan}

\author{Muneaki Hase}
\altaffiliation[Present address: ]{Institute of Applied Physics, University of Tsukuba}
\affiliation{Advanced Nano-Characterization Center, National Institute for Materials
Science, Tsukuba, 305-0047 Japan}

\author{Masahiro Kitajima}
\affiliation{Advanced Nano-Characterization Center, National Institute for Materials
Science, Tsukuba, 305-0047 Japan}

\author{Ludger Wirtz}
\affiliation{Institute for Electronics, Microelectronics, and Nanotechnology, 59652 Villeneuve d'Ascq Cedex, France}

\author{Angel Rubio}
\affiliation{European Theoretical Spectroscopy Facility, Universidad del Pa\'{i}s Vasco, Centro Mixto CSIC-UPV/EHU and DIPC, Edificio Korta, Avd. Tolosa 72, 20018 Donostia, Spain}

\author{Hrvoje Petek}
\affiliation{Department of Physics and Astronomy, University of Pittsburgh, Pittsburgh, Pennsylvania 15260, USA}

\date{\today}

\begin{abstract}
We report the ultrafast dynamics of the 47.4 THz coherent phonons of graphite interacting with a photoinduced non-equilibrium electron-hole plasma. Unlike conventional materials, upon photoexcitation the phonon frequency of graphite upshifts, and within a few picoseconds relaxes to the stationary value.  Our first-principles density functional calculations demonstrate that the phonon stiffening stems from the light-induced decoupling of the non-adiabatic electron-phonon interaction by creating the non-equilibrium electron-hole plasma. Time-resolved vibrational spectroscopy provides a window on the ultrafast non-equilibrium electron dynamics.
\end{abstract}

\maketitle


Graphite possesses highly anisotropic crystal structure, with strong covalent bonding of atoms within and weak van der Waals bonding between the hexagonal symmetry graphene sheets.  The layered lattice structure translates to a quasi-2D electronic structure, in which the electronic bands disperse linearly near the Fermi level ($E_F$) and form point-like Fermi surfaces.  The discovery of massless relativistic behavior of quasiparticles at $E_F$ of graphene and graphite has aroused great interest in the nature of carrier transport in these materials  \cite{Novoselov05,Geim07,Zhou06}.   Because of the linear dispersion of the electronic bands, the quasiparticle mass associated with the charge carrier interaction with the periodic crystalline lattice nearly vanishes, leading to extremely high electron mobilities and unusual half-integer quantum Hall effect in graphene \cite{Novoselov05,Geim07}.  Since graphite has a quasi-2D band structure very similar to that of graphene, these electronic properties may be expressed also in graphite.   

The electron-phonon (\textit{e-p}) interaction contributes to the carrier mass near $E_F$ and limits the high-field transport through the carrier scattering. The strong \textit{e-p} interaction in graphite is a distinctive characteristic of ineffective screening of the Coulomb interaction in semimetals \cite{Divin84,Spataru01}. It is expressed in the phonon frequency shift by carrier doping \cite{Dresselhaus}, electron scattering-mediated vibrational spectrum \cite{Thomsen00} and strong electronic renormalization of the phonon bands (Kohn anomalies) \cite{Piscanec04}.
Time-resolved measurements on the optically generated non-thermal electron-hole (\textit{e-h}) plasma in graphite provide evidence for the carrier thermalization within 0.5 ps both through electron-electron (\textit{e-e}) scattering and optical phonon emission \cite{Kampfrath05}.  The non-thermal carriers decay non-uniformly in phase space because of the anisotropic band structure of graphite \cite{Moos01,Spataru01}.  Quasiparticle correlations in non-thermal plasmas can also be probed from the perspective of the coherent optical phonons.  In the present study we probe the transient changes in the \textit{e-p} coupling induced by the optical perturbation of the non-adiabatic Kohn anomaly through the time-dependent complex self-energy (frequency and lifetime) of the 47 THz $E_{2g2}$ phonon of graphite.  


To probe the ultrafast response of the coherent phonons we perform transient anisotropic reflectivity measurements \cite{Hase03,Ishioka06} on a natural single crystal and highly oriented pyrolytic graphite (HOPG)  samples.  HOPG has long range order along the c-axis, but each layer consists of $\mu$m-size domains with random azimuthal orientation.  Because the phonon properties were identical, we report the results for HOPG only, whose better surface optical quality gave superior signal-to-noise ratio. 
The light source for the pump-probe reflectivity measurements is a Ti:sapphire femtosecond laser oscillator with $<$10 fs pulse duration.  The fundamental output is frequency-doubled in a $\beta$-barium borate crystal to obtain 395 nm excitation light.  The 3.14 eV photons excite vertical transitions from the valence ($\pi$) to the conduction ($\pi^*$) bands near the K point \cite{Maeda88}.  A spherical mirror brings parallel linearly polarized pump and probe beams to a common 10 ƒÊm focus on the sample with angles of 20$^\circ$ and 5$\circ$ from the surface normal, respectively.  
Pump power is varied between 5 and 50 mW (pulse fluence of 0.1 - 1mJ/cm$^2$), while probe power is kept at 2 mW.  Isotropic reflectivity change ($\Delta$R) gives a straightforward polarization dependence, while anisotropic reflectivity change ($\Delta R_{eo}=\Delta R_s - \Delta R_p$) eliminates the mostly isotropic electronic response to isloate the much weaker anisotropic contribution, which is dominated by the coherent phonon response \cite{Hase03}.  
Time delay t between the pump and probe pulses is modulated at 20 Hz to enable accumulation and averaging of up to 25,000 scans with a digital oscilloscope.  The delay scale is calibrated with recording the interference fringes of a He-Ne laser \cite{Ishioka06}.  


\begin{figure}
\includegraphics[width=8.0 cm]{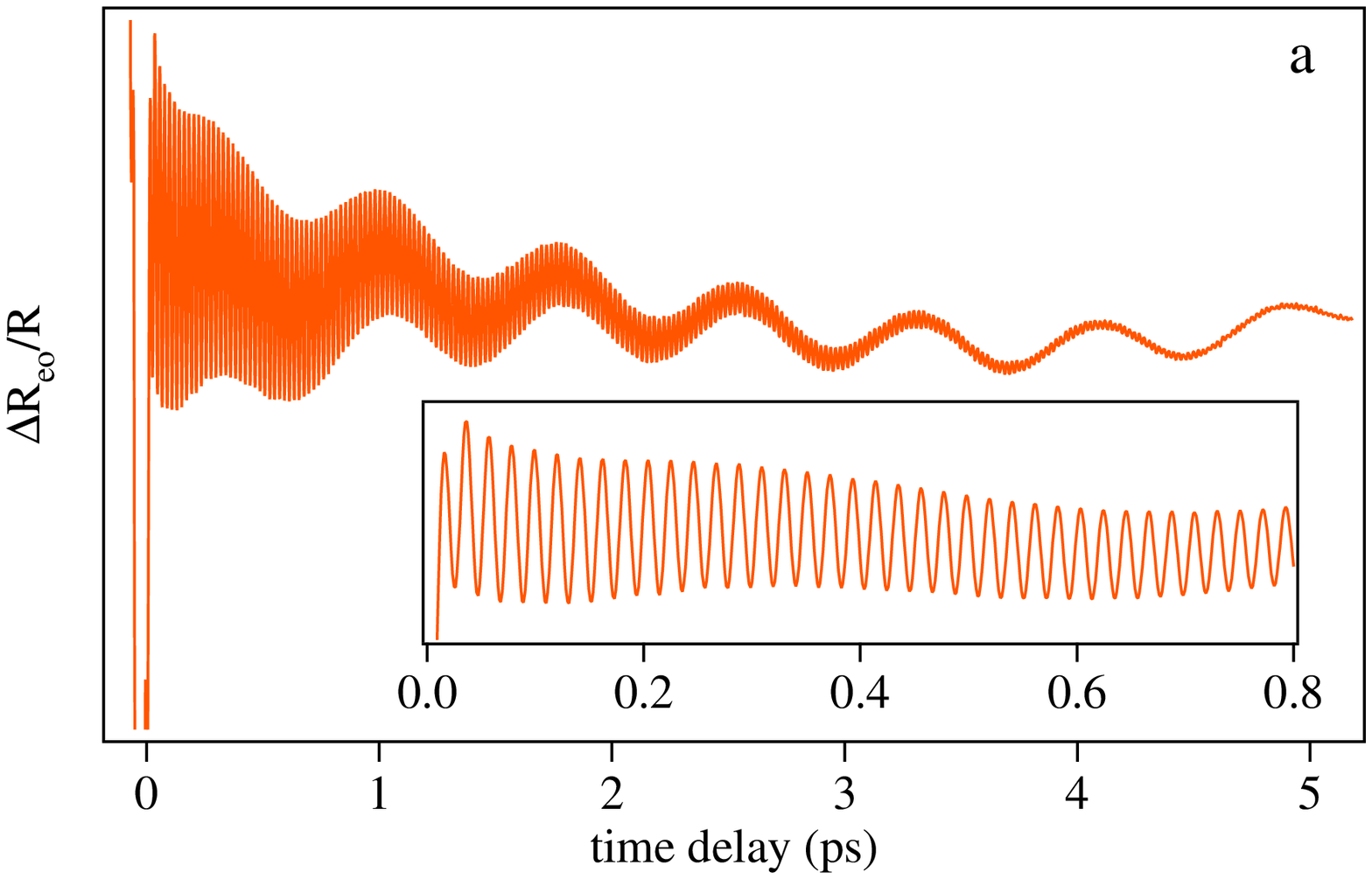}
\includegraphics[width=8.0 cm]{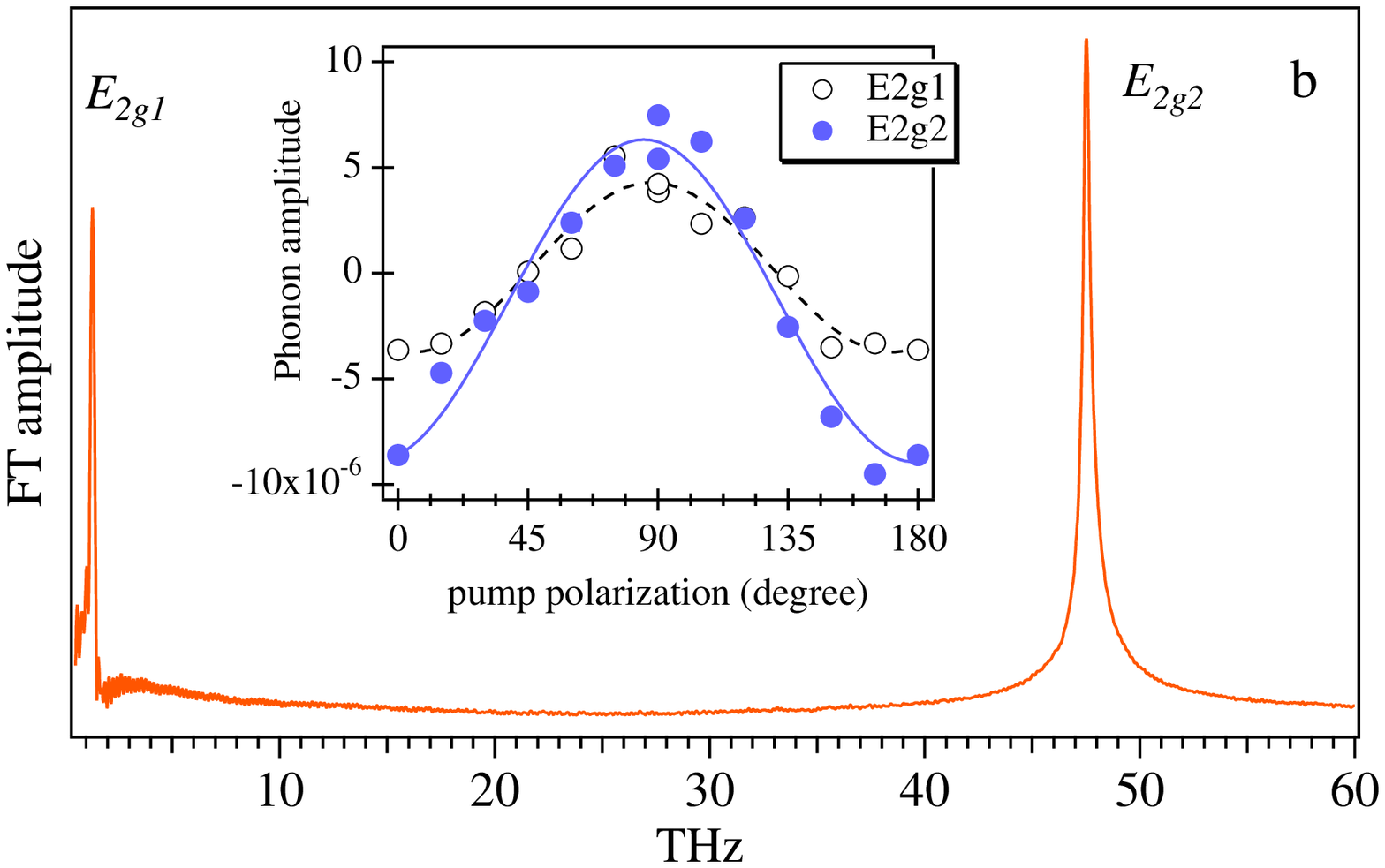}
\caption{\label{TD} (Color online) (a) Anisotropic reflectivity change $\Delta R_{eo}/R=(\Delta R_y - \Delta R_x)/R$ at pump power of 50 mW.  The inset shows an enlargement of the trace to show the high-frequency modulation. (b) FT spectrum of the time-domain trace in (a).  Inset shows the pump polarization dependence of the amplitudes of the two coherent phonons ($A_1$ and $A_2$) obtained from isotropic reflectivity ($\Delta R/R$) measurement.  The polarization angle is measured from the plane of incidence.  The probe beam is polarized at 90$^\circ$.  Solid and broken curves are fits to $\cos2\theta$ function. }
\end{figure}

Figure \ref{TD}a shows the anisotropic reflectivity change of graphite, $\Delta R_{eo}/R$,  normalized to the reflectivity without pump pulse.  After a fast and intense electronic response at t=0, the reflectivity is modulated at two disparate periods of 21 and 770 fs.  The slower coherent oscillation was previously assigned to the Raman active interlayer shear phonon ($E_{2g1}$ mode) \cite{Mishina00}.  The faster oscillation of 47.4 THz or 1580 cm$^{-1}$ is the in-plane $E_{2g2}$ carbon stretching mode \cite{Dresselhaus} corresponding to the G-peak in the Raman spectra of graphitic materials.  After decay of the electronic response, the reflectivity signal for $t>$100 fs can be fitted approximately to a sum of damped oscillations: %
$f(t)=A_1\exp(-\Gamma_1 t)\sin(2\pi\omega_1 t+\delta_1)+
A_2 \exp(-\Gamma_2 t)\sin(2\pi\omega_2 t+\delta_2)$.
The amplitudes of both phonons,  $A_1$ and $A_2$, exhibit a  $\cos2\theta$ dependence on the pump polarization angle $\theta$ with respect to the optical plane, as shown in the inset of Fig. \ref{TD}b, confirming their generation through the Raman mechanism \cite{Mishina00}.    Hereafter we focus on the previously unobserved dynamics of the fast $E_{2g2}$ phonon.

\begin{figure}
\includegraphics[width=8.0 cm]{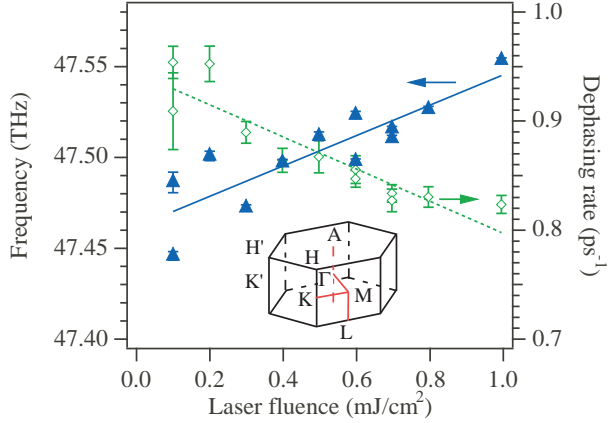}
\caption{\label{Power} (Color online) Laser fluence dependence of the dephasing rate $\Gamma_2$ and the frequency $\omega_2$ of the coherent $E_{2g2}$ phonon obtained from a fit to an exponentially damped oscillator function .  The lines are to guide the eye. Inset shows the Brillouin zone of graphite.}
\end{figure}

We measure the laser fluence dependence of the coherent phonon amplitude $A_2$, dephasing rate $\Gamma_2$, and frequency $\omega_2$  of the $E_{2g2}$ phonon that are extracted from the fit of $\Delta R_{eo}/R$ to the damped oscillator model.  The amplitude increases linearly with the fluence as expected for a $\pi-\pi^*$ transition with a single photon.  As shown in Fig. \ref{Power}, the dephasing rate  decreases as the laser fluence is increased, which is contrary to the coherent phonon response observed for other materials \cite{Hase02,Murray05,Zijlstra06}. The frequency upshift at higher fluence in Fig. \ref{Power} is equally exceptional.  Laser heating can be excluded as the origin, because the $E_{2g2}$ frequency downshifts with temperature \cite{Tan98}.  In fact, the frequency upshift under intense optical excitation has not been observed experimentally or predicted theoretically for graphite or any other solid. 

\begin{figure}
\includegraphics[width=8.0 cm]{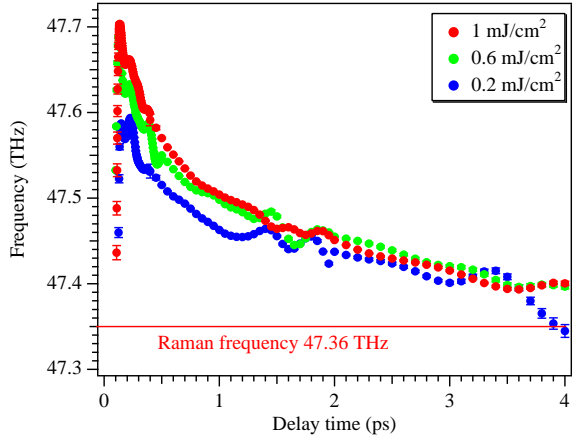}
\caption{\label{TWFT} (Color online) Time evolution of the $E_{2g2}$ phonon frequency, obtained from time-windowed FT, for different laser fluences.  The widths of the Gaussian time windows are 80 fs for $t<0.4$ ps, 300 fs for $0.4<t<2$ ps, and 800 fs for $t\geq$2 ps.  }
\end{figure}

To further characterize the unexpected frequency upshift, in Fig. \ref{TWFT} we analyze the transient reflectivity response with a time-windowed Fourier transform (FT). This analysis reveals that the phonon frequency blue-shift occurs promptly (its dynamics are obscured by the strong electronic response for delays of $<$100 fs), and recovers to its near-equilibrium value after several picoseconds.  With increasing laser fluence the initial blue-shift increases, while the asymptotic value converges on the 47.4 THz Raman frequency. The experimental phonon frequency for $t>$100 fs follows a biexponential recovery, $\omega(t)-\omega(t=\infty)=\Delta\omega_1\exp(-t/\tau_1)+\Delta\omega_2\exp(-t/\tau_2)$, with time constants of $\tau_1$=210 fs and $\tau_2$=2.1 ps, independent of excitation density.  The time scales for the recovery are in reasonable agreement with the analysis of transient terahertz spectroscopy, which gave 0.4 and 4 ps, respectively for the carrier thermalization and carrier-lattice equilibration \cite{Kampfrath05}.  The time evolution of the $E_{2g2}$ frequency implicates the interaction of coherent phonons with the photoexcited non-equilibrium carriers, as will be discussed below.


It is only recently that the observed anomalous dispersion of the high-energy phonon branches of graphite \cite{Maultzsch04} could be explained theoretically by a momentum dependent \textit{e-p} interaction (a Kohn anomaly), which leads to the renormalization (softening) of the phonon frequency \cite{Piscanec04}. Standard use of the adiabatic approximation in the previous study, however, predicted that perturbing the electronic system by electron doping would result in a downshift of phonons at the $\Gamma$ point.  Recent experiments and theoretical calculations have shown this approach to be inappropriate as the ``non-adiabatic" electronic effects, where electrons near $E_F$ cannot respond instantaneously to the lattice distortion, become important for low dimensional materials such as graphene and nanotubes \cite{Pisana07,Lazzeri06,Piscanec07}.

We perform density functional theory (DFT) calculations for a single sheet of photoexcited graphite with a new computational method that accounts for the non-adiabatic effects.  
We use DFT in the local-density approximation (LDA) as implemented in the code ABINIT \cite{Gonze02}.  Core electrons are described by Trouiller-Martins pseudopotentials and the wave-functions are expanded in plane waves with energy cutoff at 35 Hartree. For the present work the specific form of the exchange-correlation functional (LDA or GGA) does not change the emerging physical picture. For reasons of computational feasibility, we have performed calculations on single-layer graphene, as it is often done for the description of the optical phonons of graphite \cite{Maultzsch04,Wirtz04}. In order to ensure convergence of the $E_{2g2}$ phonon mode to within 0.01 THz, we use a large 61$\times$61 two-dimensional $k$-point sampling. The phonons are computed using density-functional perturbation theory \cite{Baroni01}. ``Non-adiabatic effects" are accounted for by keeping the electronic population fixed when computing the dynamical matrix. This means that the occupation of each electronic level is specified in the input of the calculation and is kept constant upon the displacement of the atoms. We neglect the effects of lattice relaxation on the phonon frequency since we checked that the effect of neutral excitation on the bond-length is very weak ($<$ 0.001 \AA) for the appropriate excitation densities. Our approach is similar to the time-dependent perturbation scheme \cite{Pisana07,Lazzeri06,Piscanec07} for the inclusion of non-adiabaticity in the combined treatment of phonons and electrons in graphite.  Furthermore, it enables us to calculate the effect of an arbitrary electron occupation far from equilibrium such as created by the vertical excitation of \textit{e-h} pairs with 3.1 eV photons.

Because the photoexcited electron distribution is time-dependent and, in principle, not known exactly, we employ three different limiting distributions.  ``As excited" distribution (AED), correspondiing to the vertical excitation of \textit{e-h} pairs with 3.1 eV photons within an energy window of $\pm$0.2 eV, simulates the distribution right after excitation with a laser pulse having a finite spectral width.  The laser fluence determines the amount of charge transferred from $\pi$ to $\pi^*$ bands.  Non-thermal distribution (NTD), in which electrons are completely depopulated in an small energy window
from top of the valence band to the bottom of conduction band, mimics the \textit{e-h} distribution after the ultrafast ($\ll$100 fs \cite{Moos01,Manzoni05}) decay of the primary excitation into the secondary \textit{e-h} pairs around $E_F$.  The width of the energy window is determined by the excited charge density.  Hot thermal distribution (TD), in which the occupation follows the Fermi-Dirac distribution with a high electronic temperature, simulates the distribution after thermalization of the electronic system ($\gtrsim$0.5 ps \cite{Kampfrath05}).  To compare with the effect of static doping reported previously \cite{Pisana07,Lazzeri06,Piscanec07}, we also present calculations with an ionized distribution (ID), in which electrons are removed from the top of the $\pi$ band.

\begin{figure}
\includegraphics[width=8 cm]{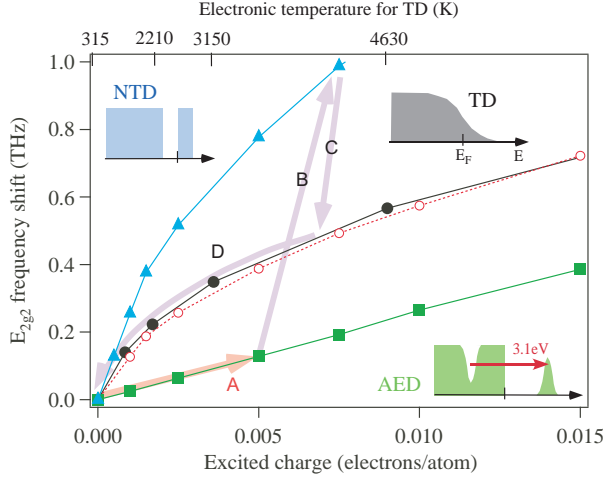}
\caption{\label{phon} (Color online) Calculated $E_{2g2}$ frequency change as a function of the excitation charge density for the as-excited distribution (AED; square), hot non-thermal distribution (NTD; triangle), thermal distribution (TD; filled circle), and ionized distribution (ID; open circle).  The top axis shows the corresponding electronic temperature $T_e$ for TD.  
Arrows show schematically the excitation and relaxation pathways for the $e-h$ distribution: quasi-instantaneous excitation by a laser pulse (A), de-excitation within $<$100 fs through the creation of secondary \textit{e-h} pairs around the Fermi level $E_F$ (B), thermalization of the \textit{e-h} plasma in $\sim$0.2 ps (C), and cooling down of the \textit{e-h} plasma in $\sim$2 ps through optical phonon emission (D).  Ultrafast phonon stiffening is ascribed to steps A and B. The highest density excitation (fluence of 1 mJ/cm$^2$) in our experiment corresponds to 5.8$\times10^{20}$ electron-hole pairs/cm$^3$ or 0.005 electrons/atom.}
\end{figure}

Figure \ref{phon} shows that all the three excited state distributions, as well as the statically doped one, lead to a stiffening of the $E_{2g2}$ phonon.  For a fixed density of the excited charge, the closer the \textit{e-h} pairs are to the $E_F$, the more pronounced is their non-adiabatic interaction with the lattice, and therefore, the stronger is their effect on the phonon stiffening.  We note that the stiffening is not accompanied by lattice deformation for the three excited distributions, contrary to the case of ID, for which the lattice both stiffens \textit{and} contracts.  The lattice stiffening for ID can be attributed to the depopulation of ƒÎ orbitals around the K and H points, which (i) suppresses the non-adiabaticity in the \textit{e-p} coupling and (ii) removes electrons with strong anti-bonding admixture \cite{Lazzeri06,Pisana07}.  Because the effect (ii) should also lead to a lattice contraction, the C-C bond stiffening under the three excited distributions is attributed to the effect (i).  This implies that the stiffening is causedd by transfer cold electrons and holes from near the $E_F$ to a hot population, which increases the ability of the electronic system to follow the ions adiabatically.  In contrast to the static doping studies \cite{Lazzeri06,Pisana07}, our observations on a neutral but non-equilibrium system address a phonon frequency shift solely of the electronic origin.

The strong dependence of the phonon stiffening on the \textit{e-h} distribution in Fig. \ref{phon} justifies interpretation of the experimental ultrafast phonon frequency changes in terms of the temporal evolution of the photoexcited \textit{e-h} plasma. The photoexcitation of carriers weakens the non-adiabatic \textit{e-p} coupling.  The reduced real part (frequency) and the increased imaginary part (decay rate) of the self-energy of \textit{e-p} interaction increases the frequency and reduces the dephasing rate of the $E_{2g2}$ mode.  The frequency recovers biexponentially on the time scales of electron thermalization and energy transfer to the lattice.  Thus, we conclude that the experimentally observed time evolution of the phonon frequency is governed by the relaxation processes of the highly non-thermal electronic population created at $t$=0 near the $K$-point (arrow A in Fig. \ref{phon}).  The very efficient \textit{e-e} scattering first brings the non-thermal \textit{e-h} carriers close to the Fermi level (near $K$ point) within a few tens of fs (arrow B), and then to electronic-thermalization in about 0.2 ps (arrow C).  This hot-thermal distribution equilibrates with the lattice through optical phonon emission on 2 ps time scale (arrow D). 


In summary, we have explored the influence of the non-equilibrium \textit{e-h} plasma on the femtosecond dynamics of the in-plane $E_{2g2}$ coherent phonon of graphite.  The time-dependent phonon frequency probes sensitively the time evolution of the transient electronic occupation distributions.  The unusual electronic stiffening of the phonon can be attributed to the excitation-induced reduction of the \textit{e-p} coupling due to quasi-2D electronic structure.  Our results offer a new paradigm of \textit{e-p} coupling, where non-equilibrium electrons impart exceptional properties to the lattice.  Similar interactions are likely to govern the \textit{e-p} coupling in related graphitic materials, such as carbon nanotubes and graphene, that are of topical interest for high-performance, nanometer scale carbon-based electronic devices.  

The authors thank O.V. Misochko for supplying single crystal graphite.  
Calculations are performed at IDRIS (project 061827), Barcelona Supercomputing Center and  UPV/EHU (SGIker Arina).  This work is supported by Kakenhi-18340093, the EU Network of Excellence Nanoquanta (NMP4-CT-2004-500198), Spanish MEC (FIS2007-65702-C02-01), French ANR, EU projects SANES (NMP4-CT-2006-017310), DNANANODEVICES (IST-2006-029192), and the NSF CHE-0650756.  H.P. thanks Donostia International Physics Center and Ikerbasque for support during the writing of this manuscript.

\bibliographystyle{apsrev}
\bibliography{E2g2}
\end{document}